\documentclass[aps,twocolumn,showpacs,preprintnumbers,amsmath,amssymb]{revtex4}
\usepackage{graphicx}
\begin{document}
%
%
\title{Magnetic friction due to vortex fluctuation}
\author{R.A. Dias}
\email{radias@fisica.ufmg.br} \affiliation{Departamento de
F\'{\i}sica - ICEX - UFMG 30123-970 Belo Horizonte - MG, Brazil}
\author{P. Z. Coura} \email{pablo@@fisica.ufjf.br}
\affiliation{Departamento de F\'{\i}sica - ICE - UFJF  Juiz de
Fora - MG, Brazil}
\author{M. Rapini} \email{mrapini@fisica.ufmg.br}
\affiliation{Departamento de F\'{\i}sica - ICEX - UFMG  30123-970
Belo Horizonte - MG, Brazil}
\author{B.V. Costa} \email{bvc@fisica.ufmg.br}
\affiliation{Laborat\'orio de Simula\c{c}\~ao - Departamento de
F\'{\i}sica - ICEX - UFMG 30123-970 Belo Horizonte - MG, Brazil}
\date{\today}
\begin{abstract}
We use Monte Carlo and molecular dynamics simulation to study a
magnetic tip-sample interaction. Our interest is to understand the
mechanism of heat dissipation when the forces involved in the system
are magnetic in essence. We consider a magnetic crystalline
substrate composed of several layers interacting magnetically with a
tip. The set is put thermally in equilibrium at temperature $T$ by
using a numerical Monte Carlo technique. By using that configuration
we study its dynamical evolution by integrating numerically the
equations of motion. Our results suggests that the heat dissipation
in this system is closed related to the appearing of vortices in the
sample.
\end{abstract}
\maketitle

%
%
\section{Introduction}
The friction between two sliding surfaces is one of the oldest
phenomena studied in natural sciences \cite{fundamentals}. In a
macroscopic scale it is known that the force of friction between
surfaces satisfies some rules: 1 - The friction is independent of
the contacting area between surfaces. 2 - It is proportional to
the normal force applied and 3 - The force of kinetic friction is
independent of relative speed between surfaces. That behavior
being the result of many microscopic interactions between atoms
and molecules of both surfaces, it is also dependent on the
roughness, temperature and energy dissipation mechanisms.
Therefore, to understand friction it is necessary to understand
its microscopic mechanisms\cite{Nanoscience}.

For several applications of now a days technology the
understanding of how heat is dissipated when mobile parts are
involved, plays an important role. The availability of refined
experimental techniques makes it now possible to investigate the
fundamental processes that contribute to the sliding friction on
an atomic scale. Issues like how energy dissipates on the
substrate, which is the main dissipation channel (electronic or
phononic) and how the phononic sliding friction coefficient
depends on the corrugation amplitude were addressed , and
partially solved, by some groups. \cite{smith,liebsch}. Less known
is the effect on friction of a magnetic tip moving relative to a
magnetic surface. Applications of sub-micron magnets in
spintronics, quantum computing, and data storage demand a huge
understanding of the sub-micron behavior of magnetic materials.
The construction of magnetic devices has to deal with distances of
nanometers between the reading head and the storage device. That
makes the study of tribological phenomena crucial to understand
and produce technologically competitive devices
\cite{BoLiu,Suh,Bhushan}. In particular the dissipation of heat in
magnetic dispositives is a very serious problem. For example, in a
magnetic hard disk for data storage the reading head passing close
to the surface of the disk transfers momentum to it. That momentum
transference rises locally the temperature. Depending on the rate
transfer and the capability of the disk to transfer that energy to
the neighborhood some, or all, the information stored in the disk
can be lost. In the last decade, the progress in the magnetic
recording media and the reading head technology has made the
recording density doubled almost every two years. The magnetic bit
size used in the most advanced hard-disk-drives is as small as
$0.5 \times 0.05 m^2$, while a giant magneto-resistance head is
used to read the bit. This magnetic bit size can still be
diminished by using materials with high magnetic anisotropy.
However, the paramagnetic limit, when the anisotropy energy
becomes comparable to the thermal fluctuations, is not far to be
attained. That is a physical limitation to the technology for the
production of magnetic recording media. In that case, the
comprehension of the microscopic mechanism of heat dissipation is
crucial. Since friction is an out-of-equilibrium phenomenon its
study presents numerous theoretical and experimental challenges.

\begin{figure}
\includegraphics[width=4.5cm]{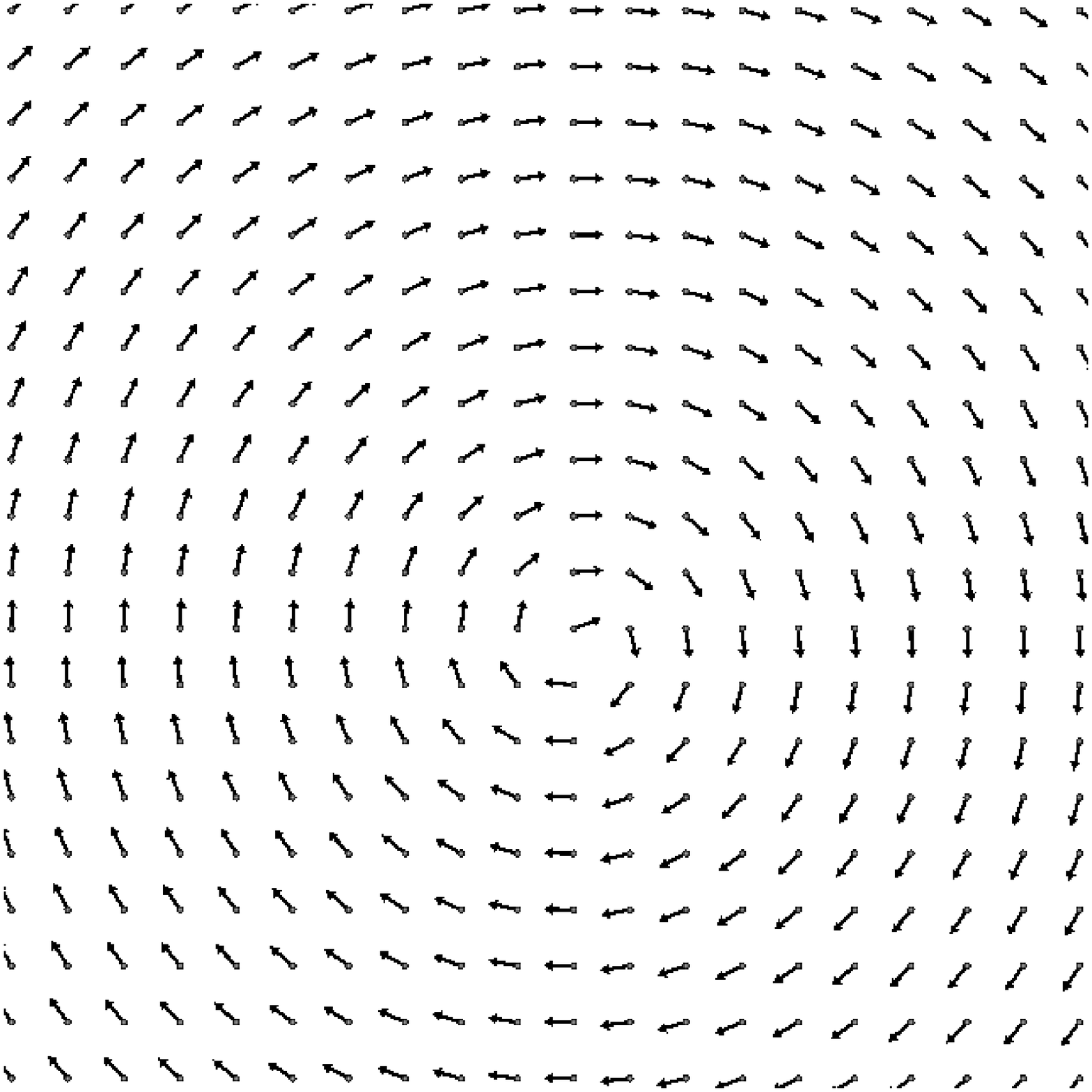}~~~~
\includegraphics[width=4.5cm]{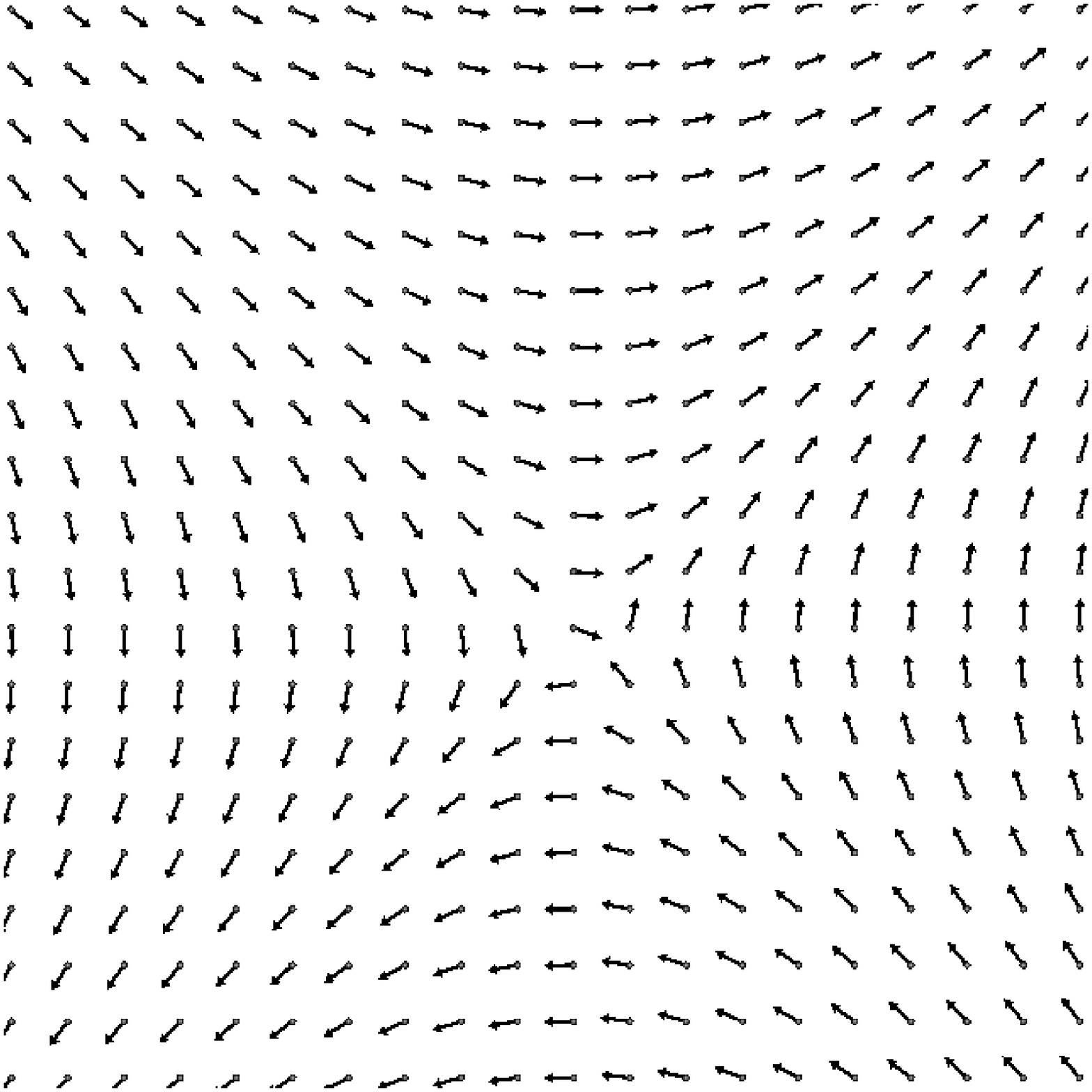}\\
\caption{Schematic view of a vortex (left) and anti-vortex (right)
for spins of the same length.} \label{vortex}
\end{figure}

In magnetic films for which the exchange interaction is less than
the separation between layers form quasi 2 dimensional (2d)
magnetic planar structures. In general the magnetization of such
films is confined to the plane due to shape anisotropy. An
exception to that is the appearing of vortices in the system. A
vortex being a topological excitation in which the integral of
line of the field on a closed path around the excitation core
precess by $2 \pi$ ($-2 \pi$) (See figure \ref{vortex}.). To the
purpose of avoiding the high energetic cost of non-aligned
moments, the vortices develop a three dimensional structure by
turning out of the plane the magnetic moment components in the
vortex core.\cite{evaristo-bvc} For data storage purposes,
magnetic vortices are of high interest since its study provides
fundamental insight in the mesoscopic magnetic structures of the
system\cite{Choe}.

In this paper we use a combined Monte Carlo-Molecular Dynamics
(MC-MD) simulation to study the energy dissipation mechanism in a
prototype model consisting of a reading head moving close to a
magnetic disk surface. A schematic view is shown in figure
\ref{esquemaMagFric}. Our model consists of a magnetic tip (The
reading head) which moves close to a magnetic surface (The disk
surface.). The tip is simulated as a cubic arrangement of magnetic
dipoles and the surface is represented as a monolayer of magnetic
dipoles distributed in a square lattice. We suppose that the
dipole interactions are shielded, so that, we do not have to
consider them as long range interactions. This trick simplifies
enormously the calculations putting the cpu time inside reasonable
borders. The dipole can be represented by classical spin like
variables ${\vec S} = S_x \hat{x} + S_y \hat{y} + S_z \hat{z}$.
The total energy of this arrangement is a sum of exchange energy,
anisotropy energy and the kinetic energy due to the relative
movement between the tip and the surface as follows.
%
%
\begin{equation}\label{hamiltonian}
H = \sum_{i=1}^{N_h} \frac{\vec p_{(h-s)i}^2}{2m_{(h)i}} +
    U_{spin} + U_{h-s},
\end{equation}
where $U_{spin} = U_h + U_s$.
\begin{eqnarray}\label{potencial}
U_h &=& -\frac{J_h}{2}\sum_{<i,j>} \left( S_{hi}^x  \cdot S_{hj}^x +
                                      S_{hi}^y  \cdot S_{hj}^y +
                                      \lambda_h S_{hi}^z  \cdot S_{hj}^z  \right)   \nonumber \\
                                  &-&    D_h\sum_{i=1}^{N_h}\left( S_{hi}^z  \right)^2  \\
U_s &=& -\frac{J_s}{2}\sum_{<i,j>} \left( S_{si}^x  \cdot S_{sj}^x +
                                      S_{si}^y  \cdot S_{sj}^y +
                                      \lambda_s S_{si}^z  \cdot S_{sj}^z  \right)   \nonumber \\
                                  &-&    D_s\sum_{i=1}^{N_s}\left( S_{si}^z  \right)^2
\end{eqnarray}
and
\begin{equation}\label{interacao}
U_{h-s} = -\sum_{i,j} J_{h-s}\left( \left| \vec{r}_{hi} -
\vec{r}_{sj} \right| \right) (\vec S_{hi}  \cdot \vec S_{sj})
\end{equation}
with
\begin{equation}\label{J-corte}
J_{h-s}\left( \left| \vec{r}_{hi}  - \vec{r}_{sj} \right| \right) =
J_0 \exp\{-\alpha \left( \vec r_{hi-sj} - r_0  \right)^2\}
\end{equation}
In equation \ref{hamiltonian} the first term, ${\vec
p_{(h-s)i}^2}/{2m_{(h)i}}$  , stands for the relative kinetic
energy : surface-reading head (s-h). The second term, $U_{spin}$,
accounts for the magnetic dipoles interactions: in the tip ($U_h$)
and in the surface ($U_s$). The last term, $U_{h-s}$, is the
interaction energy between the tip and the surface. The symbol
$<ij>$ means that the sums are to be performed over the first
neighbors. For the tip-surface interaction, we suppose that the
coupling, $J_{h-s}$, is ferromagnetic. By considering that
$J_{h-s}$ is a function of distance, will allow us to study the
effects of the relative tip-surface movement. The exchange
anisotropic term $\lambda$, controls the kind of vortex which is
more stable in the lattice.
%
%
There is a critical value of the anisotropy, $\lambda_c \approx
0.7J$ \cite{evaristo-bvc}, such that for $\lambda < \lambda_c$ the
spins inside the vortex core minimizes the vortex energy by laying
in an in-plane configuration. For $\lambda
> \lambda_c$ the configuration that minimizes the vortex energy is for
the spins close to the center of the vortex to develop a large
out-of-plane component. The site anisotropy, $D$, controls the
out-of-plane magnetization of the model. It is well known that a
quasi-two dimensional system with interaction as in equation
\ref{potencial} undergoes a $BKT$ phase transition for $D_h$
sufficiently small (In general $D/J << \lambda$.), at some
critical temperature $T_{BKT}$. If, $D/J = \lambda$ = 0, $T_{BKT}
\approx 0.7$. For $D/J >> \lambda$ the system has a second order
phase transition at $T_c$ which depends on $D$.

In the sections below we will discuss the importance of vortex
formation for the energy dissipation of a magnetic material. In
the section \ref{section2} we introduce general aspects of the
numerical method used, in the section \ref{section3} we discuss
our results and in section \ref{section4} we present our
conclusions.

\section{Simulation Background\label{section2}}

In this section we describe the numerical approach we have used to
simulate the model. The simulation is done by using a combined
Monte Carlo-Molecular Dynamics (MC-MD) procedure. The particles in
our simulation move according Newton's law of motion which can be
obtained by using the hamiltonian \ref{hamiltonian}. The  spins
evolve according to the equation of motion \cite{gerling}
\begin{equation}\label{Sponto}
\frac{d \vec{S}_{hi,si}}{dt}=-\vec{S}_{hi,si}\times \frac{\partial
H}{\partial\vec{S}_{hi,si}}.
\end{equation}
As we are mainly interested in the magnetic effects we consider
the particles as fixed letting the tip slid over the surface. This
generates a set of $3N+3$ coupled equations of motion which are
solved by increasing forward in time the physical state of the
system in small time steps of size $\delta t=10^{-3} J^{-1}$. The resulting
equations are solved by using Runge-Kutta's method of
integration\cite{Allen,Rapaport,Berendsen}.
%
\begin{figure}
\includegraphics[width=7.0cm]{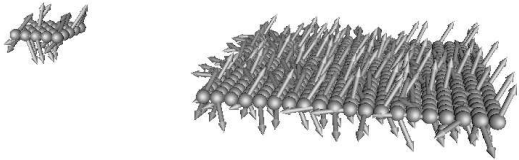}\\
\caption{Schematic view of our simulation arrangement. The tip is
simulated as a $5 \times 5$ rigid lattice of particles each
carrying a spin $\vec S$. The surface is simulated as an
arrangement of size $20 \times 20$. The tip is allowed to slid
over the surface with initial velocity $v = v_0 \hat{x}$. Periodic
boundary conditions are used for the in plane directions of the
surface. The arrows represent the spins directions.}
\label{esquemaMagFric}
\end{figure}
%
The surface is arranged as a rigid $20a \times 20a$ square lattice
with periodic boundary conditions in the $xy$ direction, where $a$
is the lattice spacing. The head is simulated as a rigid $5a
\times 5a$ square lattice. With no loss of generality the lattice
spacing will be taken as $a=1$ from now on. In figure
\ref{esquemaMagFric} we show a schematic view of the arrangement
used in our simulation. Initially we put the tip at a large
distance from the surface, so that, $J_{h-s}=0$. By using the MC
approach we equilibrate the system at a given temperature, $T$. By
controlling the energy of the system we consider that the system
is in equilibrium after $10^5$ MC steps. Once the thermal
equilibrium is reached we observe the evolution of the separated
systems for a small time interval, $\tau$, for posterior
comparison. After that an initial velocity, $v = v_0 \hat{x}$ is
given to the tip. We follow the system's evolution by storing all
positions, velocities and spin components at each time step for
posterior analysis. A quantity of paramount importance for us is
the vortex density, $\varphi_v(t)$, at the surface layer. We
calculate the vortex density as a function of temperature and time
for several values of the parameters $\lambda_s$, $\lambda_h$ and
$D_s$, $D_h$. As discussed in the following the vortex density
will be related to energy dissipation. The energy is measured in
units of $ J_h = J_s = J$, temperature in units of $J/k_B$, time
in units of $J^{-1}$ and velocity in units of $a/J^{-1}$ where
$k_B$ is the Boltzmann constant.
\section{Results\label{section3}}
The system is simulated for several temperatures and anisotropy
parameters. In all of them we have fixed $J_h = J_s = J$, $J_0 = 2
J$ and $\alpha=1$. In the first set of simulations we use
$\lambda_s = \lambda_h = 1 J$  and the site anisotropy $D_s = D_h
= 0.1 J$, $-0.1 J$ to obtain a system with out-of-plane
(Ising-like) and in-plane ($XY$) symmetries respectively. In the
second set, we put $D_s = D_h = 0$ and $\lambda_s = \lambda_h =
0.6 J,0.9 J$ to get vortices with in-plane and out-of-plane cores
respectively.
%
\begin{figure}[htbp]\vspace{0.5cm}
\includegraphics[width=8.0cm]{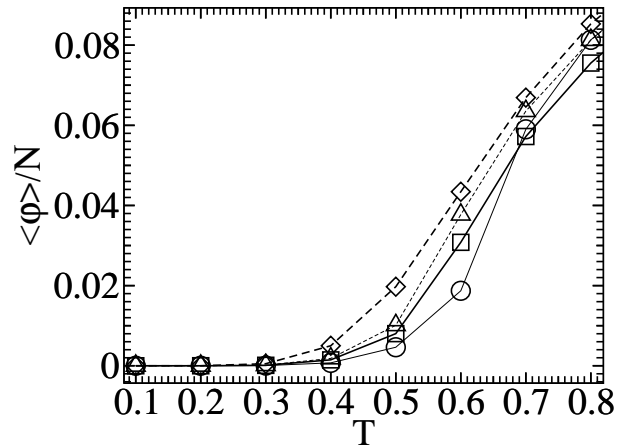}\\
\caption{Vortex Density as a function of the temperature. The
symbols are for $\lambda_s = \lambda_h = 1$ with $D_h=D_s=0.1J$
($\circ$), $D_h=D_s=-0.1J$ ($\square$) and $D_s = D_h = 0$ with
$\lambda_s=\lambda_s=0.6J$ ($\Diamond$) and
$\lambda_h=\lambda_s=0.9J$ ($\vartriangle$). The lines are only
guides to the eyes.} \label{DenVortxT-4Sim}
\end{figure}
%
Before we start the time evolution, we have to know how the vortex
density, $\varphi_v$, depends on the temperature. In figure
\ref{DenVortxT-4Sim} we show a plot of $\varphi_v$ as a function
of the temperature for the simulated models. The vortex density
increases monotonically with $T$. If there is any relation between
vortices and energy dissipation, it is natural to think that an
increase in the vortex density is related to an increase in
temperature.

With the head far from the surface, we start the time evolution of
the system at $t=0$ with $v=0$. This part of the simulation serves
as a guide to the rest of the simulation. Only thermal
fluctuations of the vortex density can be seen. At $t = 200$ the
tip is released with initial velocity $v_0$. For $t > 200$, the
reading head starts to interact with the surface. Some kinetic
energy is transferred from the head to the surface and we expect
the vortex density to increase. We will see in the following that
depending on the initial conditions and the symmetry of the system
(See equation \ref{hamiltonian}) several things can happen.
\begin{figure}[htbp]
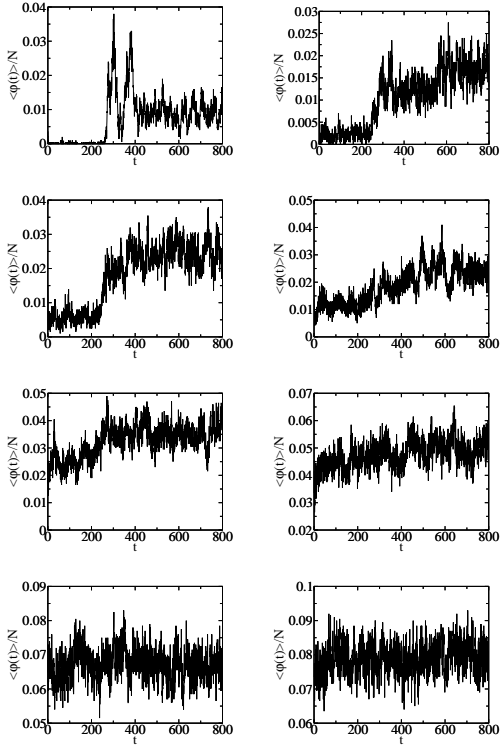
\vspace{0.25cm}
\includegraphics[width=3.0cm]{fig.3a.eps}~~~~
\includegraphics[width=3.0cm]{fig.3b.eps}\\
\vspace{0.35cm}
\includegraphics[width=3.0cm]{fig.3c.eps}~~~~
\includegraphics[width=3.0cm]{fig.3d.eps}\\
\vspace{0.35cm}
\includegraphics[width=3.0cm]{fig.3e.eps}~~~~
\includegraphics[width=3.0cm]{fig.3f.eps}\\
\vspace{0.35cm}
\includegraphics[width=3.0cm]{fig.3g.eps}~~~~
\includegraphics[width=3.0cm]{fig.3h.eps}\\
\caption{Vortex Density as a function of time for $D_h=D_s=0.1J$
and $\lambda_s=\lambda_s=1J$. The graphics from left to right and
top to bottom correspond to the temperatures $T=0.1, 0.2, 0.3,
0.4, 0.5, 0.6$. The system has an out-of-plane symmetry.}
\label{DenVortxTimeD01}
\end{figure}
\begin{figure}[htbp]
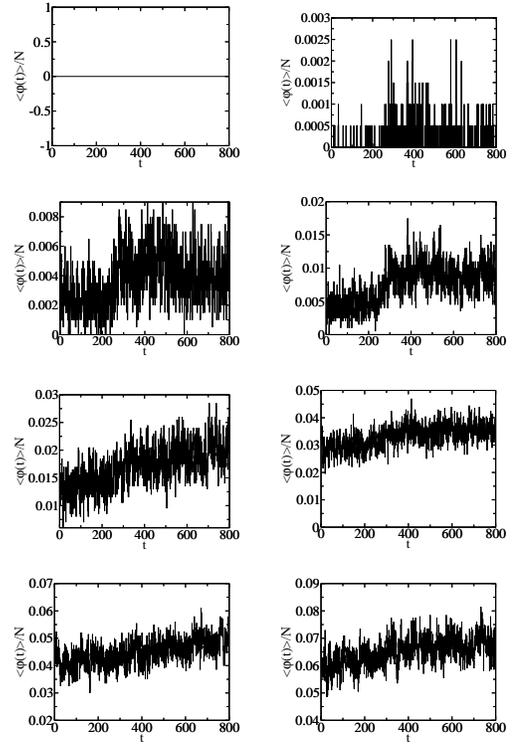
\vspace{0.25cm}
\includegraphics[width=3.0cm]{fig.4a.eps}~~~~
\includegraphics[width=3.0cm]{fig.4b.eps}\\
\vspace{0.35cm}
\includegraphics[width=3.0cm]{fig.4c.eps}~~~~
\includegraphics[width=3.0cm]{fig.4d.eps}\\
\vspace{0.35cm}
\includegraphics[width=3.0cm]{fig.4e.eps}~~~~
\includegraphics[width=3.0cm]{fig.4f.eps}\\
\vspace{0.35cm}
\includegraphics[width=3.0cm]{fig.4g.eps}~~~~
\includegraphics[width=3.0cm]{fig.4h.eps}\\
\caption{Vortex Density as a function of time for $D_h=D_s=-0.1J$
and $\lambda_s=\lambda_s=1J$. Temperatures are the same as in
figure \ref{DenVortxTimeD01}. The system has an in-plane
symmetry.} \label{DenVortxTimeD-01}
\end{figure}
\begin{figure}[htbp]\vspace{0.25cm}
\includegraphics[width=3.0cm]{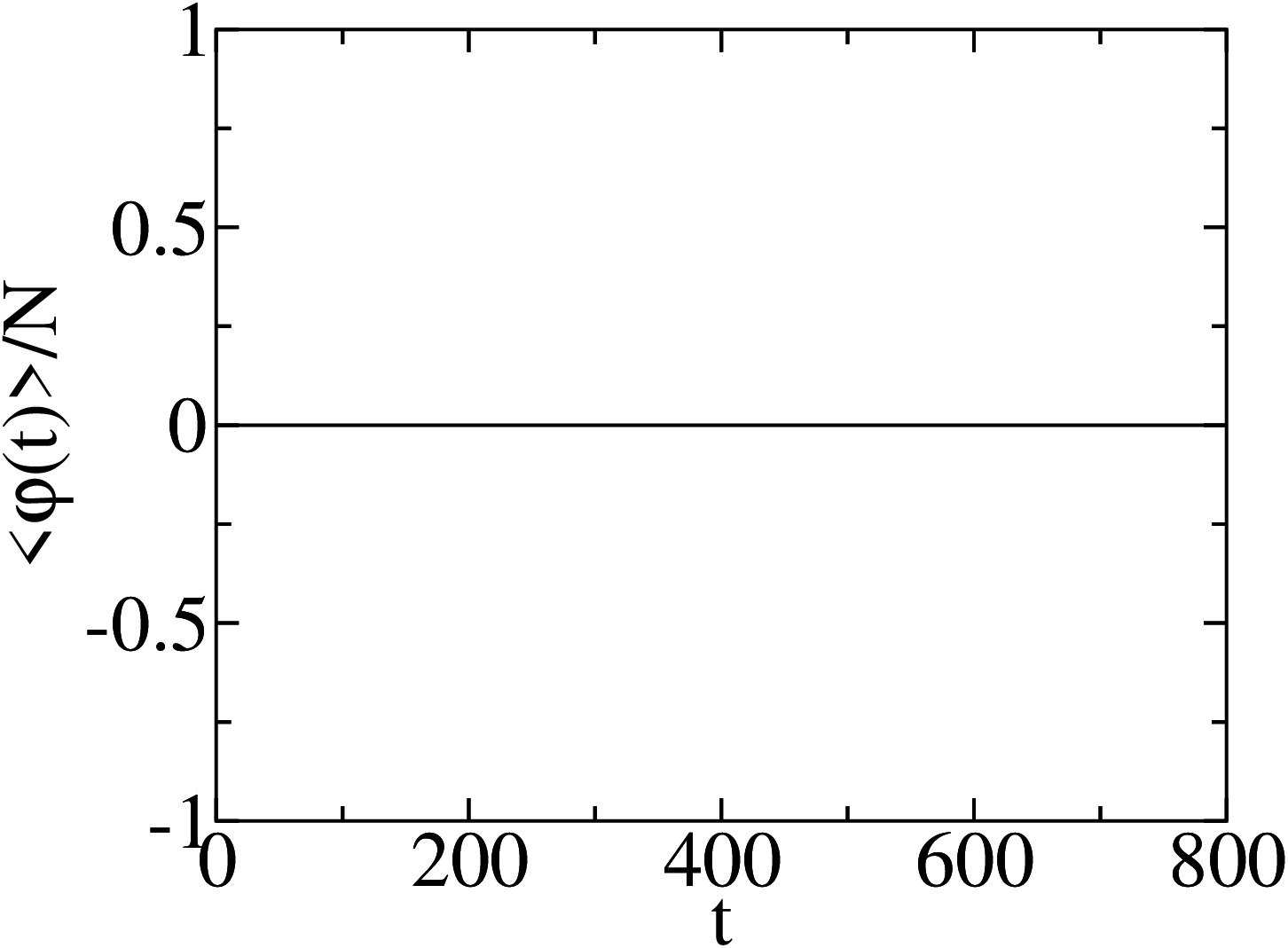}~~~~
\includegraphics[width=3.0cm]{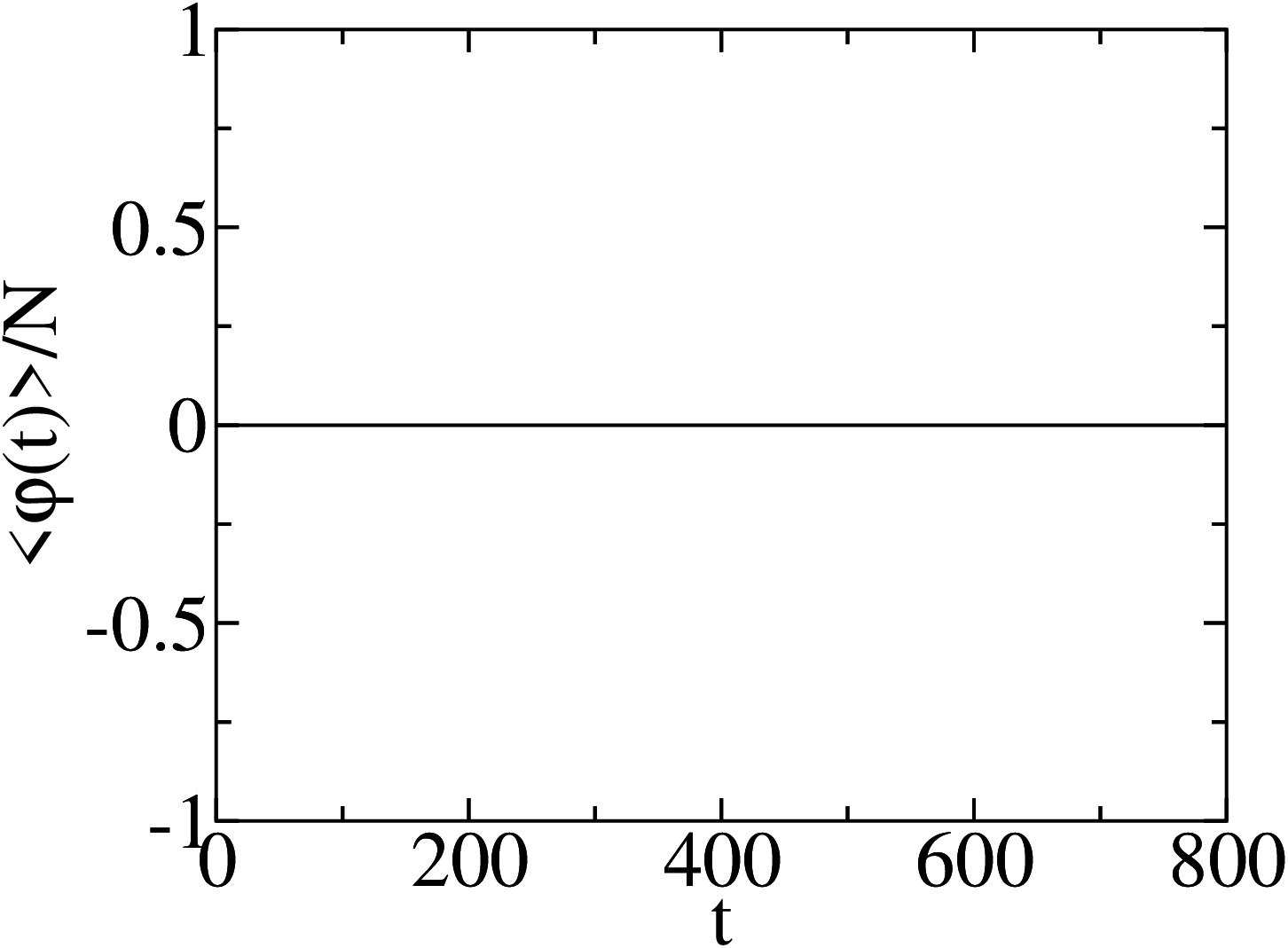}\\
\vspace{0.35cm}
\includegraphics[width=3.0cm]{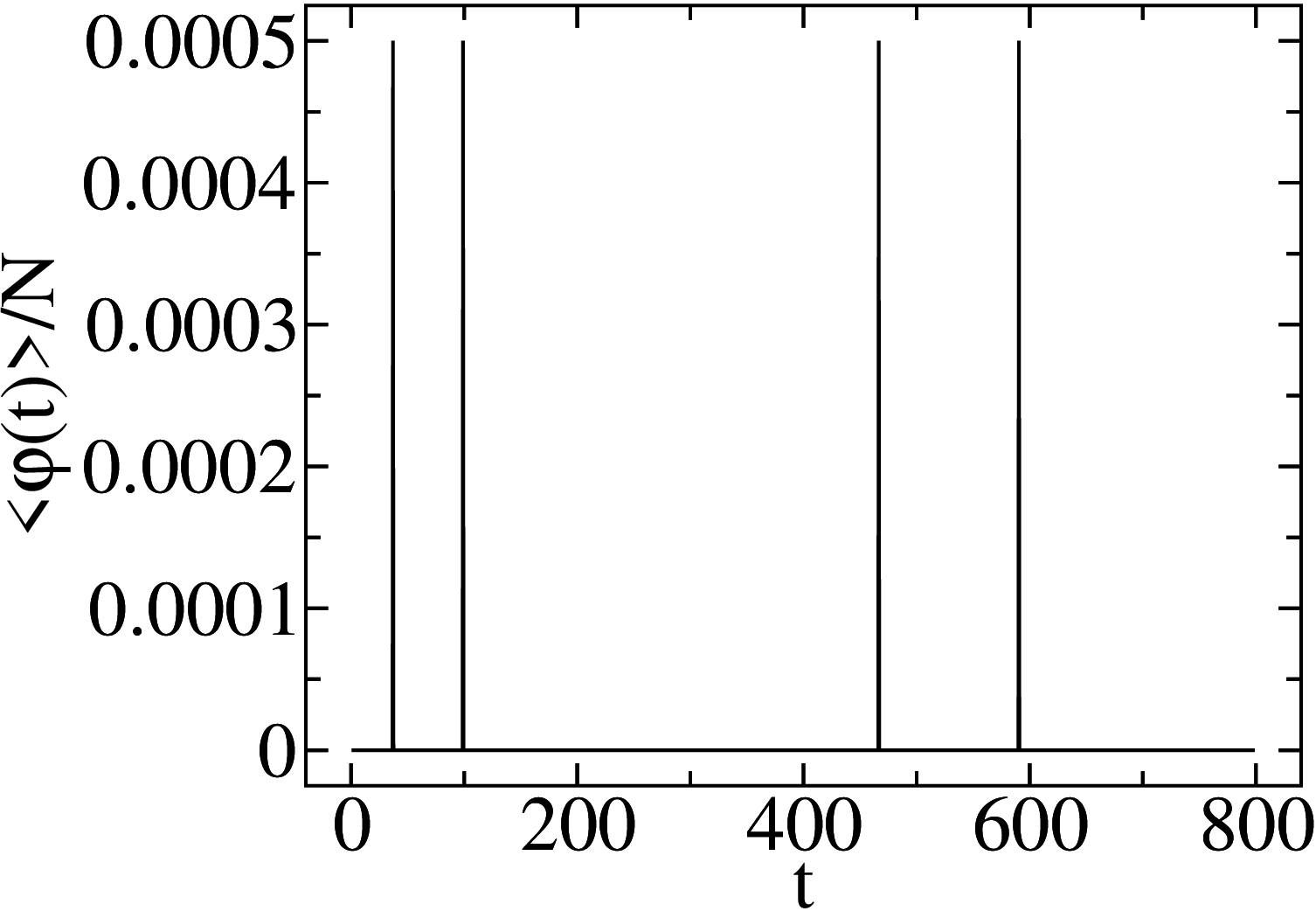}~~~~
\includegraphics[width=3.0cm]{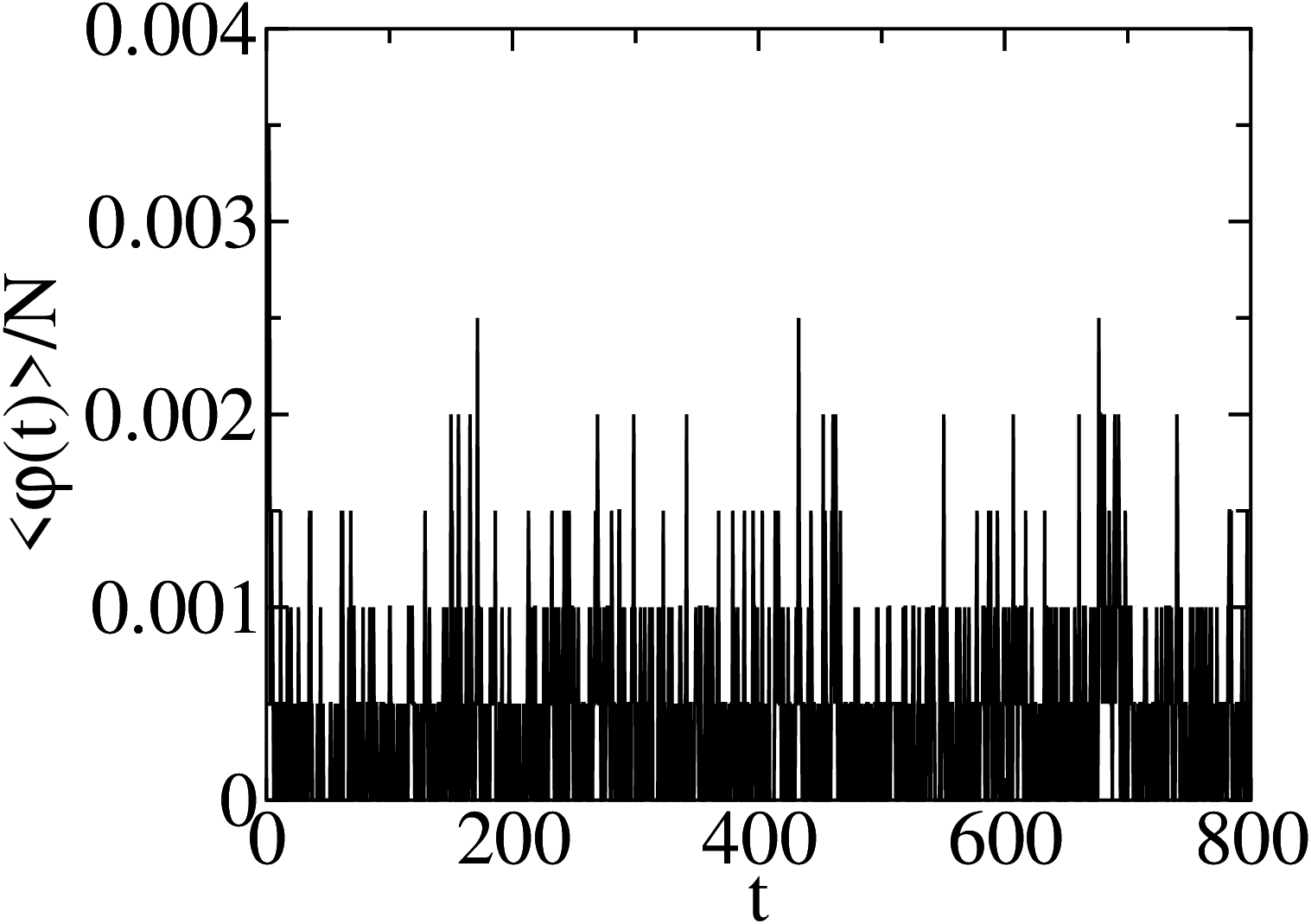}\\
\vspace{0.35cm}
\includegraphics[width=3.0cm]{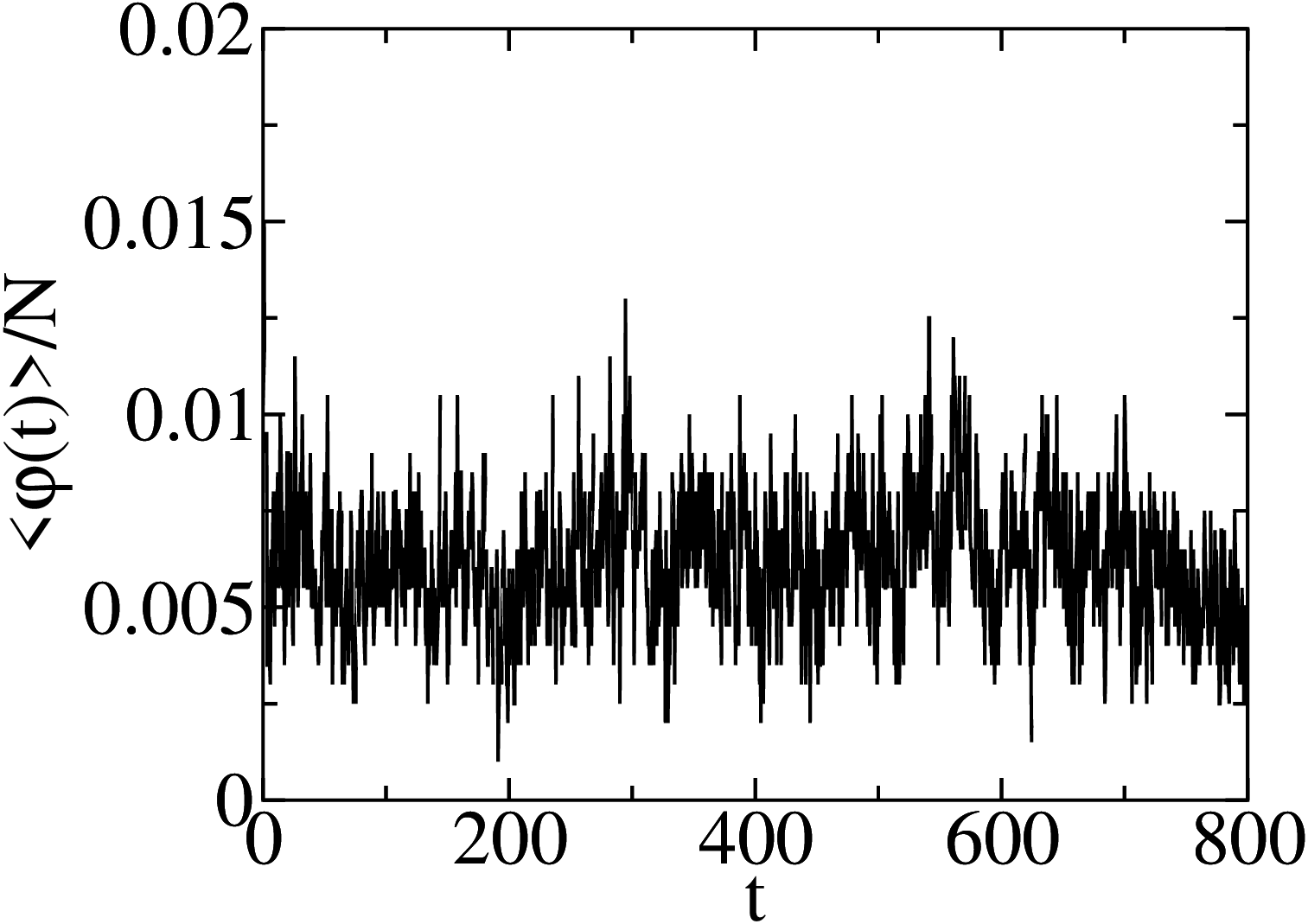}~~~~
\includegraphics[width=3.0cm]{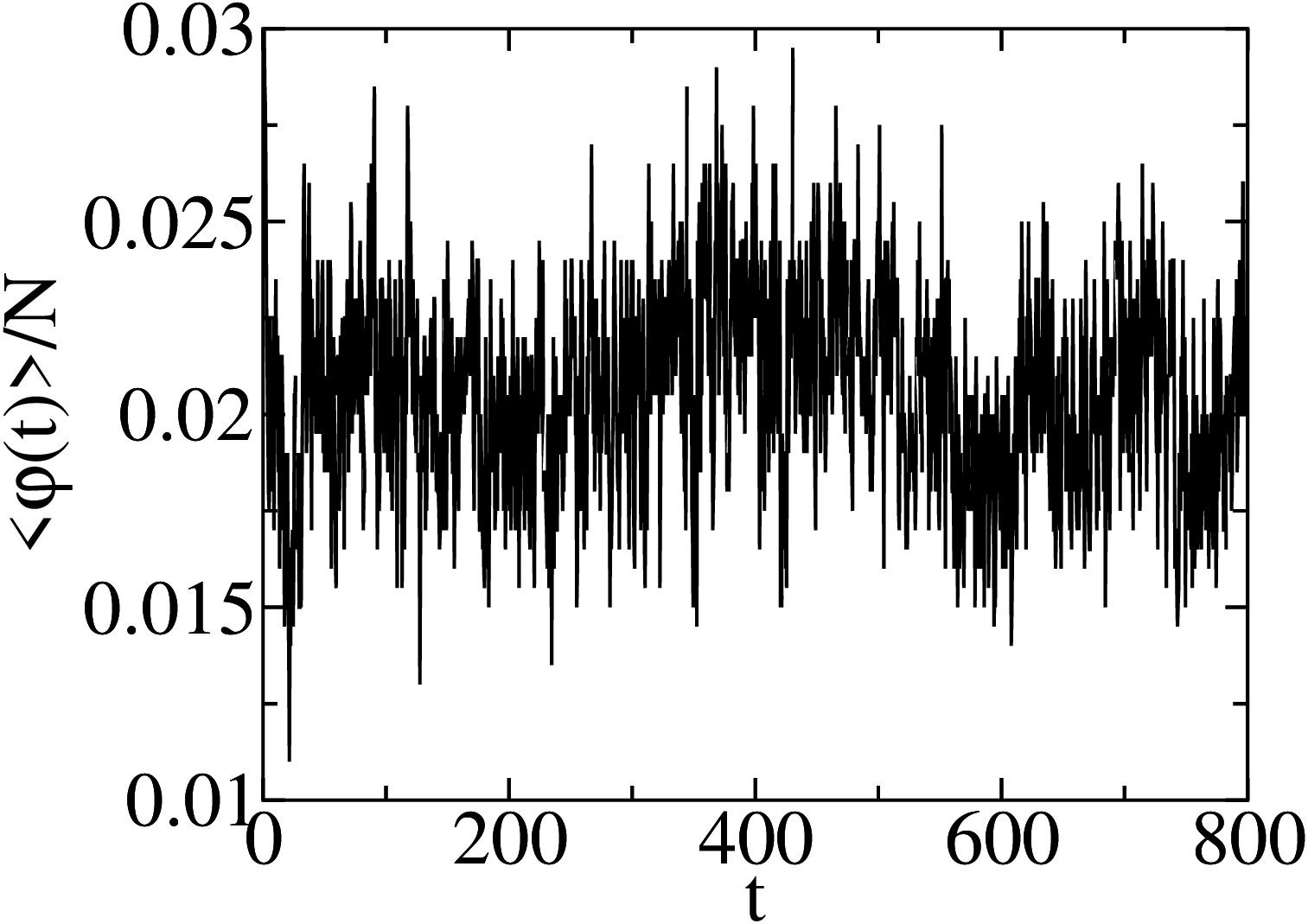}\\
\vspace{0.35cm}
\includegraphics[width=3.0cm]{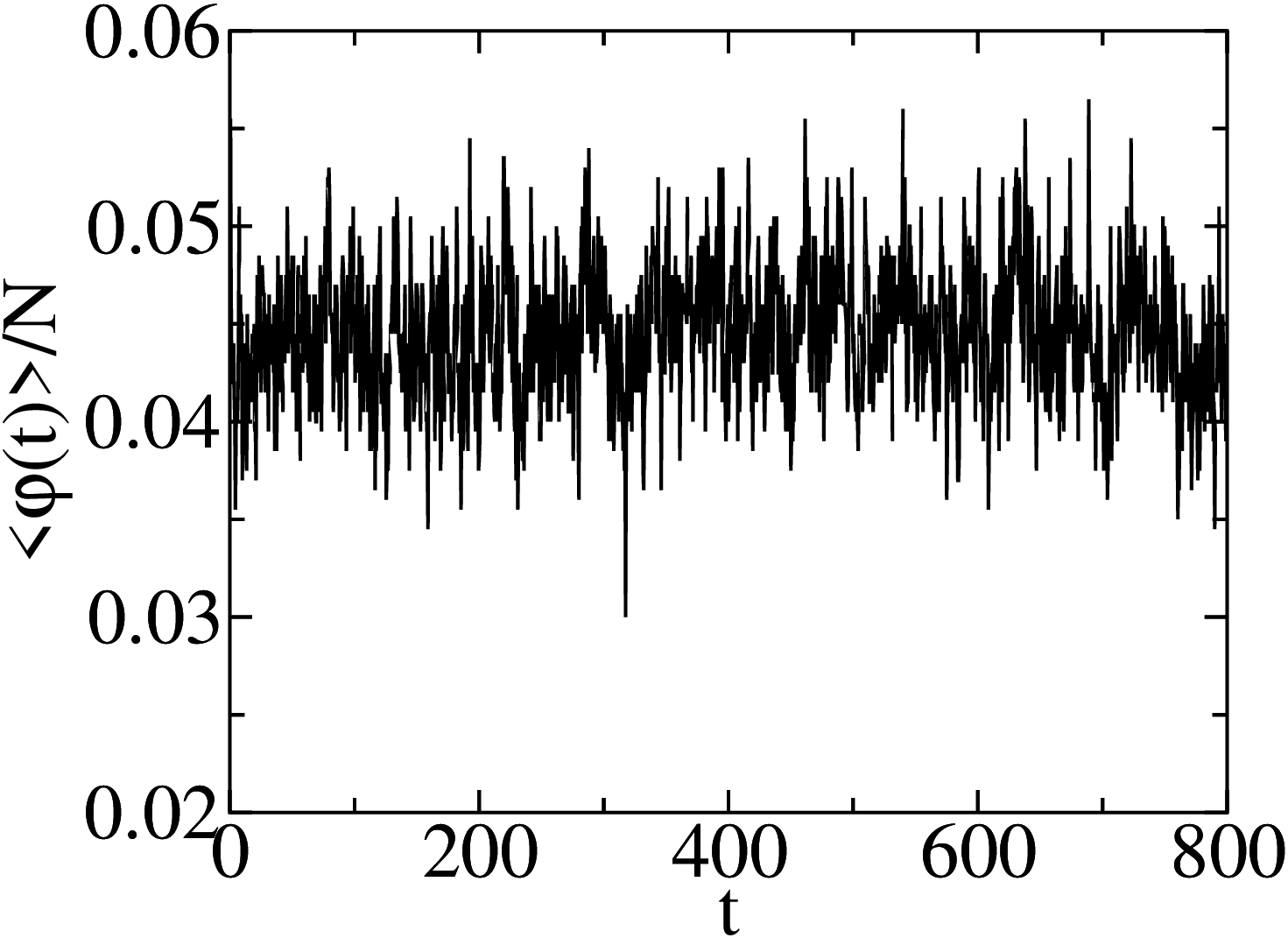}~~~~
\includegraphics[width=3.0cm]{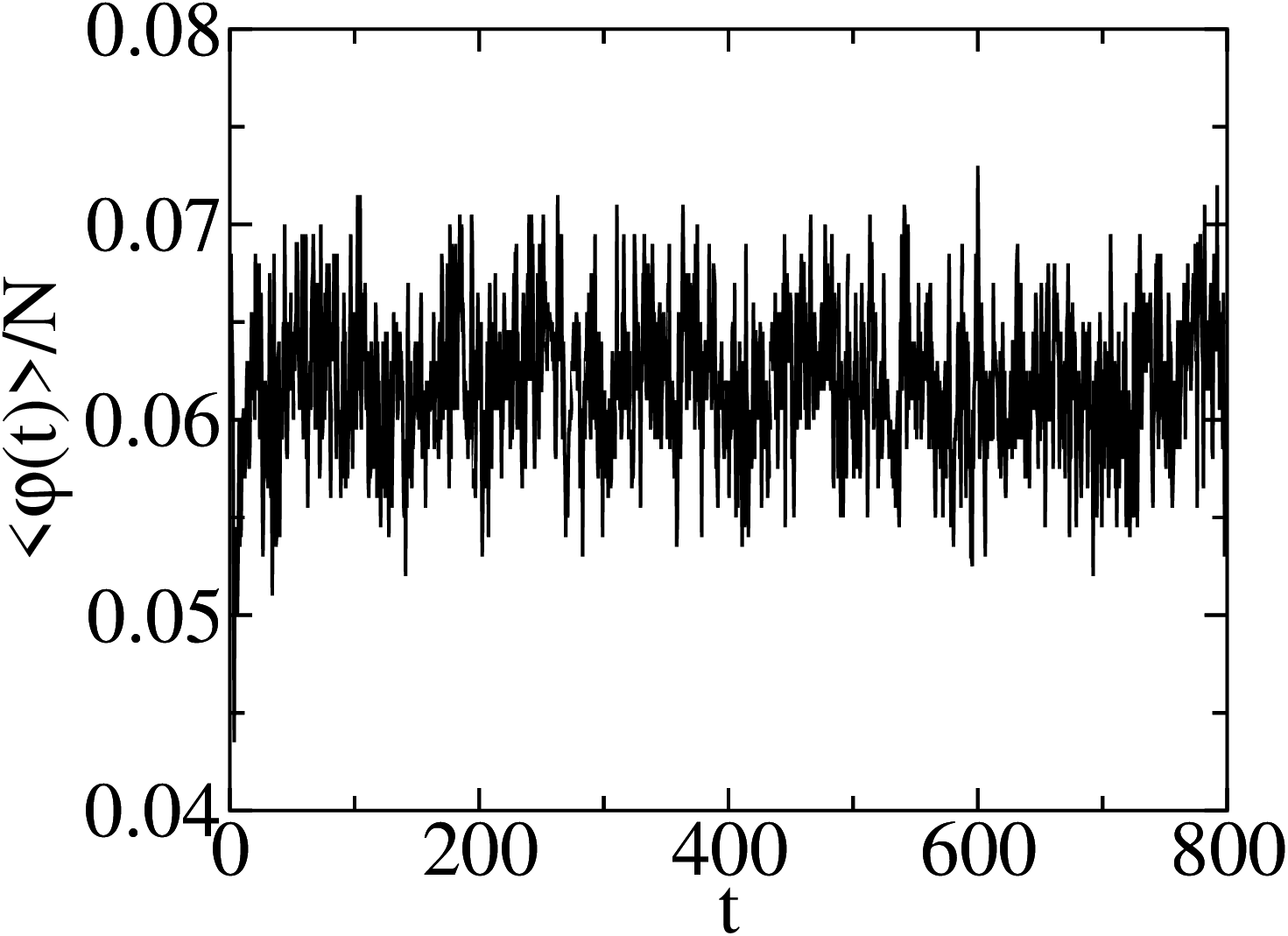}\\
\caption{Vortex Density as a function of time for
$\lambda_s=\lambda_s=0.6J$ and $D_h=D_s=0$. Temperatures are the
same as in figure \ref{DenVortxTimeD01}. The vortex has an
in-plane structure.} \label{DenVortxTimeL06}
\end{figure}
\begin{figure}[htbp]\vspace{0.25cm}
\includegraphics[width=3.0cm]{fig.6a.eps}~~~~
\includegraphics[width=3.0cm]{fig.6b.eps}\\
\vspace{0.35cm}
\includegraphics[width=3.0cm]{fig.6c.eps}~~~~
\includegraphics[width=3.0cm]{fig.6d.eps}\\
\vspace{0.35cm}
\includegraphics[width=3.0cm]{fig.6e.eps}~~~~
\includegraphics[width=3.0cm]{fig.6f.eps}\\
\vspace{0.35cm}
\includegraphics[width=3.0cm]{fig.6g.eps}~~~~
\includegraphics[width=3.0cm]{fig.6h.eps}\\
\caption{Vortex Density as a function of time for
$\lambda_s=\lambda_s=0.9J$ and $D_h=D_s=0$. Temperatures are the
same as in figure \ref{DenVortxTimeD01}. The vortex has an
out-of-plane structure.} \label{DenVortxTimeL09}
\end{figure}
\begin{figure}[htbp]
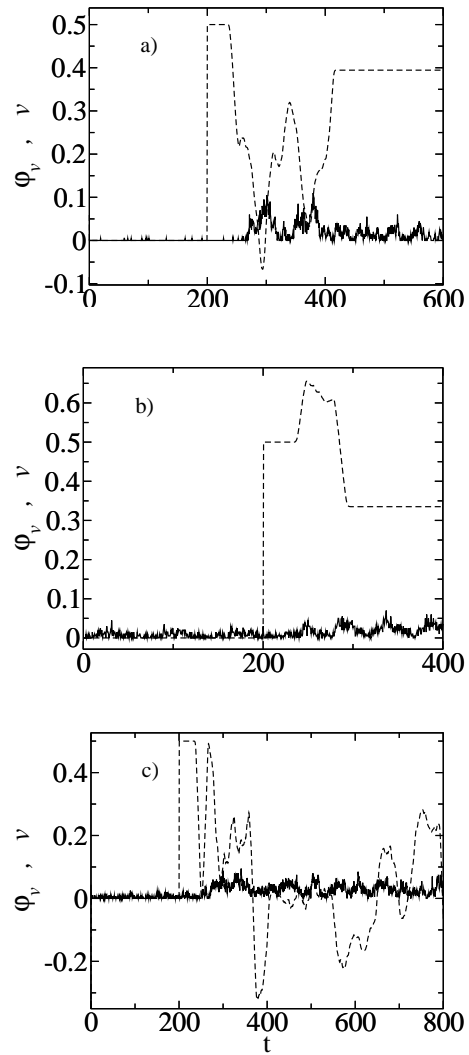

\vspace{0.25cm}
\includegraphics[width=6cm]{fig.7a.eps}\\
\vspace{0.45cm}
\includegraphics[width=6cm]{fig.7b.eps}\\
\vspace{0.45cm}
\includegraphics[width=6cm]{fig.7c.eps} \\
\vspace{0.35cm} \caption{ Instantaneous velocity and vortex
density as a function of time for $D_h=D_s=0.1J$ and
$\lambda_s=\lambda_s=1J$. The full line is the vortex density and
the dashed line is for the instantaneous velocity. From top to
bottom $T=0.1,0.2$. The initial velocity is $v=0.5$ in all cases.
The vortex density is multiplied by a factor $2$ as a matter of
clarity.} \label{DEcMed}
\end{figure}
%
\begin{figure}[htbp]
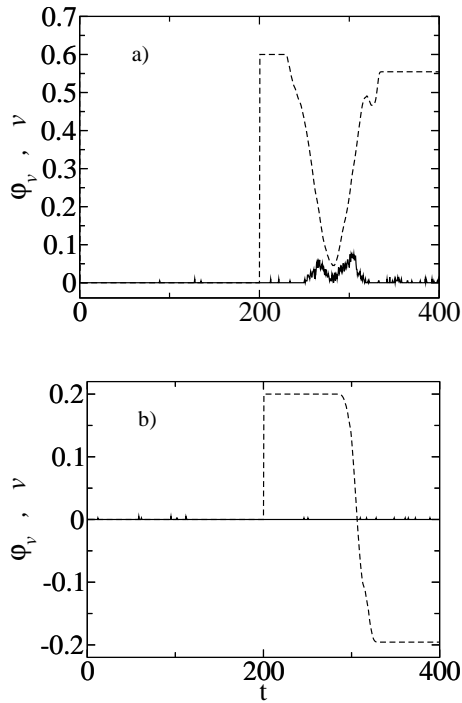

\vspace{0.25cm}
\includegraphics[width=6cm]{fig.8a.eps}\\
\vspace{0.45cm}
\includegraphics[width=6cm]{fig.8b.eps}\\
\vspace{0.35cm} \caption{Instantaneous velocity and vortex density
as a function of time for $D_h=D_s=0.1J$ and
$\lambda_s=\lambda_s=1J$. The full line is the vortex density and
the dashed line is for the instantaneous velocity. The temperature
is $T=0.2$ in both cases. From top to bottom the initial velocity
is $v=0.6,0.2$. The vortex density is multiplied by a factor $2$
as a matter of clarity.}
 \label{fig4}
\end{figure}
In the figures \ref{DenVortxTimeD01} to \ref{DenVortxTimeL09} we
show plots of the averaged vortex density as a function of the
time for the set of the simulated parameters as discussed above.
In each figure, the graphics from left to right and top to bottom
correspond to $T=0.1, 0.2, 0.3, 0.4, 0.5, 0.6, 0.7$ and $0.8$.

For the system with out-of-plane symmetry (Figure
\ref{DenVortxTimeD01}) we observe that for low temperature the
vortex density augments when the tip passes over the surface.
Initially the vortex density grows reaching quickly a constant
average. For higher temperature the vortex density is almost
insensitive to the tip indicating that the energy transfer becomes
more difficult. At low temperature it is easier to excite the
vortex mode since they have low creation energy due to the
out-of-plane spin component. At higher temperature the system is
already saturated and creating a new excitation demands more
energy. For the in-plane symmetry (Figure \ref{DenVortxTimeD-01})
the situation is opposite. At low temperature there is no energy
transfer to vortex modes. Creating a vortex demands much energy
because the spin component of the vortex is almost whole in-plane.
At higher temperature the system is soft so that it can absorb
energy augmenting the vortex density. Eventually it reaches
saturation at high enough temperature. For the case when the
vortex has an in-plane-symmetry ($\lambda < \lambda_c$), shown in
figure \ref{DenVortxTimeL06}, the average vortex density is
constant even at higher temperatures. For ($\lambda > \lambda_c$)
(Figure \ref{DenVortxTimeL09}) the situation is similar to that
where the system has a global out-of-plane symmetry (Compare to
figure \ref{DenVortxTimeD01}).

In figure \ref{DEcMed} we show the velocity of the head and the
vortex density as a function of time plotted in the same graphic
for some interesting situations. The vortex density is multiplied
by a factor of $2$ as a matter of clarity. In the first plot we
observe that the kinetic energy of the tip is transferred to the
surface. The head stops, moves back and forth and escapes from the
surface influence. At higher temperatures the situation is a bit
more complex. Depending on the initial condition the head passes
through the surface region just augmenting the vortex density
(\ref{DEcMed}.b) or it can be trapped in the surface region, as
seen in figure \ref{DEcMed}.c. The cost for the increase in the
vortex density is a decrease in the kinetic energy.

In figure \ref{fig4} $a$ and $b$ we show the velocity and vortex
density as a function of time for two different initial velocities
($v = 0.6,0.2$ respectively), at the same temperature $T = 0.2$.
For the higher velocity, the tip decreases its velocity almost up
to stop, however, its kinetic energy is sufficient to go across
the surface region. For a lower velocity, figure \ref{fig4}.$b$,
the tip collides elastically with the surface. Because there is no
lost in kinetic energy the vortex density is conserved. We note
that the increase in the vortex density it is not an instantaneous
response to the diminishing of the kinetic energy of the tip. It
may be due to an intermediate mechanism: The kinetic energy is
used to excite spin waves in the surface. Because there is no
mechanism of energy dissipation, part of the energy contained in
the spin waves is transferred to vortex excitations.
%
\section{Conclusions\label{section4}}
    We have used Monte Carlo and spin dynamics simulation to study the
interaction between two magnetic mobil parts: a magnetic reading
head dislocating close to a magnetic surface. Our interest was to
understand the mechanism of heat dissipation when the forces
involved in the system are magnetic in essence. To simulate the
surface we have considered a magnetic crystalline substrate
interacting magnetically with a magnetic tip. From the results
presented above we have strong evidences that vortices play an
important role in the energy dissipation mechanism in magnetic
surfaces. The augmenting of the vortex density excitations in the
system increases its entropy. That phenomenon can blur any
information eventually stored in magnetic structures in the
surface. An interesting result is the velocity behavior of the tip
passing close to the surface. In principle we should expect that
the velocity will always diminish, as an effect of the interaction
with the surface. However, for certain initial conditions the
effect is opposite. The tip can oscillate, be trapped over the
surface or even be repelled. In the case of an elastic collision
the vortex density remains unchanged. If the vortex density
increases the tip's kinetic energy diminishes. However, the
increase in the vortex density is not an instantaneous response to
the diminishing of the kinetic energy of the tip. We suspect that
an intermediate mechanism involving spin wave excitations is
present intermediating the phenomenon.

The effects on friction observed in our simulations demonstrate
that when pure magnetic forces are involved they are quite
different from ordinary friction. There are two points that should
be interesting to study: The effect of normal forces applied to
the system and how the observed effects depend on the contacting
area between surfaces.
\section{Acknowledgments}
Work partially supported by CNPq (Brazilian agencies). Numerical
work was done in the LINUX parallel cluster at the Laborat\'{o}rio
de Simula\c{c}\~{a}o Departamento de F\'{i}sica - UFMG.
%
%

%
\end{document}